\documentclass[prd,aps,floats,onecolumn,twoside,preprintnumbers,
superscriptaddress,floatfix,nofootinbib,showpacs]{revtex4-2}
\usepackage{graphicx,colordvi,pstricks,rotating,amsmath}
\usepackage{amssymb,epstopdf,bbm}
\usepackage{slashed}
\usepackage[T1]{fontenc}

\usepackage{simplewick}
\makeatletter  
\newcommand{\fmslash}[2][0mu]{ 
   
    {\fmsl@sh\scriptscriptstyle{#1}{#2}}}
\newcommand{\fmsl@sh}[3]{%
\m@th\ooalign{$\hfil#1\mkern#2/\hfil$\crcr$#1#3$}}

\begin{document}

\title{Seiberg-Witten maps and scattering amplitudes of  noncommutative QED}

\author{Josip Trampeti\'{c}}
\affiliation{Ru\dj er Bo\v{s}kovi\'{c} Institute, Division of Experimental Physics, Bijeni\v{c}ka 54, 10000 Zagreb, Croatia}
\email{josip.trampetic@irb.hr}
\affiliation{Max-Planck-Institut f\"ur Physik, (Werner-Heisenberg-Institut), F\"ohringer Ring 6, 
D-80805 M\"unchen, Germany}
\email{trampeti@mppmu.mpg.de}
\author{Jiangyang You}
\affiliation{Ru\dj er Bo\v{s}kovi\'{c} Institute, Division of Physical Chemistry, Bijeni\v{c}ka 54, 10000 Zagreb, Croatia}
\email{jiangyang.you@irb.hr}

\newcommand{\tr}{\hbox{tr}}
\def\BOX{\mathord{\vbox{\hrule\hbox{\vrule\hskip 3pt\vbox{\vskip
3pt\vskip 3pt}\hskip 3pt\vrule}\hrule}\hskip 1pt}}

\date{today}

\begin{abstract} 
The connection between tree-level scattering amplitudes and the Seiberg-Witten (SW) map in the Moyal deformed U(1) noncommutataive quantum electrodynamics (NCQED) is studied. We show that in the minimal U(1) NCQED based on a reversible Seiberg-Witten (SW) map, SW map induced interactions cancel each other in all tree-level scattering amplitudes and leave them identical to the Moyal NCQED without SW map. On the other hand, the two-by-two Compton and light-by-light  scattering amplitudes deviate from minimal model when irreversible SW map is used. Therefore the reversibility of SW map and equivalence between NCQED before and after SW map manifest as an identity between the tree-level scattering amplitudes.
\end{abstract}


\maketitle    

\section{Introduction}
It is long known that a noncommutative (NC) $\rm U_\star(N)$ gauge theories can be converted into a $\rm U(N)$ gauge theories by Seiberg-Witten (SW) map, as required by the symmetry of the underline D-brane systems~\cite{Seiberg:1999vs}. Originally the SW map was constructed as a reversible map between the noncommutative $\rm U_\star(N)$ and commutative $\rm U(N)$ gauge fields. Soon after it was realized that an irreversible SW map from commutative gauge fields valued in any Lie algebra to deformed/NC gauge fields valued in the universal enveloping algebra of the Lie algebra can also be realized, resulting in many more Moyal-deformed classical gauge theories \cite{Madore:2000en,Jurco:2000fb,Jurco:2001rq,Jurco:2001kp,arXiv0711.2965B,arXiv0909.4259B,Calmet:2001na,Aschieri:2002mc,Barnich:2002pb,Martin:2010ng,Horvat:2011qn,Martin:2012aw,Horvat:2012vn,Dimitrijevic:2012pk,Aschieri:2014xka,Dimitrijevic:2014iwa,Trampetic:2015zma}. SW maps have also been used for constructing gauge theories on other noncommutative spaces~\cite{Dimitrijevic:2005xw,Dimitrijevic:2011jg,Chaichian:2021egm}.

Since 1999 till today the noncommutative  gauge field theories have been studied intensively as perturbative quantum field theories~\cite{Gomis:2000zz,Aharony:2000gz,Minwalla:1999px,Hayakawa:1999yt,Matusis:2000jf,VanRaamsdonk:2000rr,VanRaamsdonk:2001jd,Schupp:2008fs,Horvat:2011bs,Martin:2020ddo,Martin:2016zon,Martin:1999aq}. Nowadays, a number of open questions still remain. One frequently raised issue is how the theories based on a SW map, either reversible or irreversible, could be connected to the Moyal NC $\rm U_\star(N)$ gauge theories without SW map. In two prior works, we addressed this issue by showing that the on shell effective actions of noncommutative U(1) gauge theories before and after a reversible SW map, obtained by the background field method, are formally equivalent/dual to each other  \cite{Martin:2016hji,Martin:2016saw}. The equivalence relation is verified by computing directly the one-loop two-point function before and after SW map.

Recently we have shown that the scattering amplitudes of all tree-level two-by-two scattering processes in a minimal U(1) noncommutative quantum electrodynamics (NCQED) model based on a reversible SW map are identical to the same theory without SW map~\cite{Latas:2020nji}. We view this result as a strong suggestion that the aforementioned equivalence relation would be realized in tree-level scattering amplitudes as an identity.  It then becomes, apparently, interesting to investigate whether this identity can be generalized to all tree-level scattering amplitudes of the same model and thus manifest the desired equivalence relation. Another, related question, is what kind of changes one would expect from a NCQED model based on irreversible SW map, in which the formal equivalence relation no longer holds.

In this article we address both issues with explicit results. We first demonstrate that all tree-level scattering amplitudes of the U(1) noncommutative QED minimal/first model based on a reversible SW map are identical to the corresponding scattering amplitudes of the NC U$_\star$(1) model without a SW map by a constructive method. Using the relations obtained when proving the general identity, we proceed to show that the same identity is lost already in the two-by-two fermion-photon and photon-photon scattering processes in the NCQED second model containing two differently charged fermions which is based on an irreversible SW map. Therefore, the identical tree-level scattering amplitudes in the U(1) NCQED model(s) are indeed the consequence of the reversibility of the SW map, as we have expected from the general formal equivalence.

This article is organized as follows:\\
In Sec. II we defined the minimal/first model, as the noncommutative Moyal deformed QED action induced by means of a $\theta$-exact reversible Seiberg-Witten  map for a single gauge field (photon) and single charged fermion.  
In Sec. III we deal with amplitudes in the gauge sectors of NCQED, i.e., with a diagrammatic approach to the photon scattering amplitudes in the U(1) SW mapped gauge theory. In particular in Secs. IIIA and IIIB we consider the 
$\gamma\gamma\to\gamma\gamma$ scattering amplitude and their generalization to the to $n$-photon scattering amplitudes, respectively. Section IIIC is devoted to the scattering amplitudes including fermions. In Sec. IV we consider and analyse the second model of the deformed QED built by a $\theta$-exact irreversible Seiberg-Witten  map, with two gauge sectors and with fermion sector containing two fermions $\psi_1$ and $\psi_2$, with two different charges \cite{Horvat:2011qn,Horvat:2012vn}, and compared it with the minimal/first model. Finally, in Sec. V we present a discussion and the conclusion.

\section{The minimal Moyal deformed NCQED model based on reversible SW map}

We consider the minimal/first NCQED model in terms of one noncommutative U(1) gauge field and one left-charged fermion field, $A_\mu$ and $\Psi$, respectively~\cite{Madore:2000en},
\begin{eqnarray}
^1S_{\rm min}&=&\int-\frac{1}{4e^2}\;F_{\mu\nu}[{A}_\mu(e a_\mu)]\star F^{\mu\nu}[{A}_\mu(e a_\mu)]\;+\;\bar\Psi(\psi, e a_\mu)\star(i\slashed{D}-m)\Psi(\psi, e a_\mu),
\label{NCminAction}\\
F_{\mu\nu}&=&\partial_\mu A_\nu - \partial_\nu A_\mu -i[A_\mu\stackrel{\star}{,}A_\nu],\;D_\mu\Psi=\partial_\mu\Psi - i A_\mu\star\Psi.
\nonumber
\end{eqnarray}
The connection between the NC and commutative quantities is given by the reversible SW map \cite{Seiberg:1999vs}, which is expressed exactly with respect to the NC deformation parameter 
$\theta^{\mu\nu}$ as an expansion over coupling constant $e$/formal power of fields~\cite{Schupp:2008fs,Martin:2012aw,Trampetic:2015zma}. 

\section{Scattering amplitudes of the minimal model}

We start scattering amplitude analysis in minimal/first model with pure gauge sector (and deal with the inclusion of fermions later on) by recording property of the $\theta$-exact SW mapped U(1) gauge theory,  all photon self-interaction vertices ($V$) satisfy the following identity, 
\begin{equation}
k^{\mu_i}_i\cdot V_{\mu_1...\mu_i...\mu_n}(k_1,...,k_i,...,k_n)=0,\;\; i=1...n.
\label{property}
\end{equation}
\begin{figure*}[t]
\begin{center}
\includegraphics[width=10cm,angle=0]{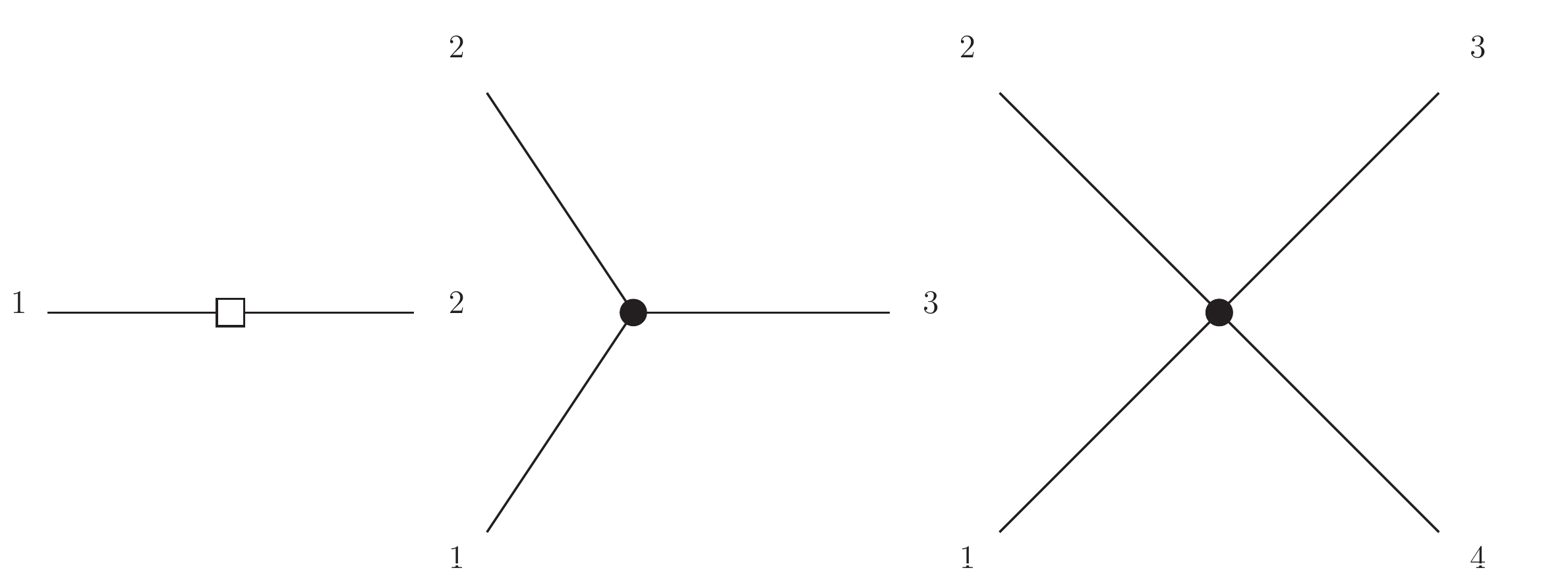}
\end{center}
\caption{The backbone of interacting structures}
\label{backbone}
\end{figure*}
Now we consider the $n\ge 3$-photon self-interaction terms. There are two origins of them:
\begin{figure*}
\begin{center}
\includegraphics[width=8cm,angle=0]{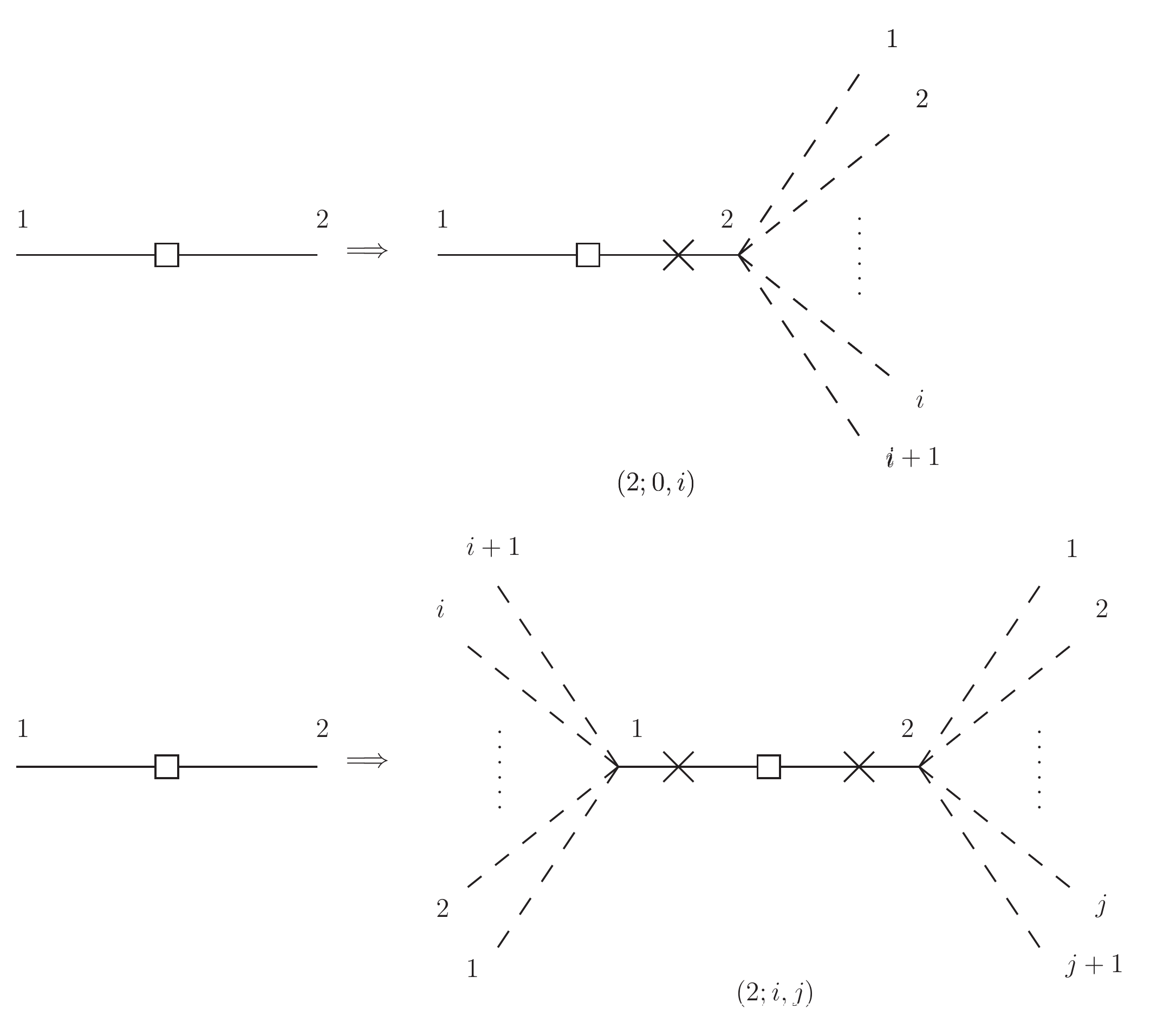}
\end{center}
\caption{The $(2,0,i)$ and $(2; i, j)$ type interactions. Here full lines denote photon legs which are at 
the zeroth order of SW map/directly from the unmapped backbone schematic diagrams in Fig.~\ref{backbone}, while dashed lines represents photon legs generated by the SW map. Crosses denote legs of the backbone interaction which are replaced by SW-mapped composite field operators.}
\label{2ij}
\end{figure*}

\noindent
a. First, we have the three-photon and the four-photon vertices from the $\rm U_\star(1)$ NCYM without SW map, which generate the $\gamma\gamma\to\gamma\gamma$ scattering amplitude \cite{Hewett:2000zp,Latas:2020nji,Horvat:2020ycy}.
\\
b. Second,  the SW mapped terms originated from the quadratic (denoted as a box with two free lines attached), 
cubic (black dot with three free lines attached) and quartic terms (black dot with four free lines attached) from the 
SW unmapped actions, shown schematically in Fig.~\ref{backbone}.

\begin{figure*}
\begin{center}
\includegraphics[width=8cm,angle=0]{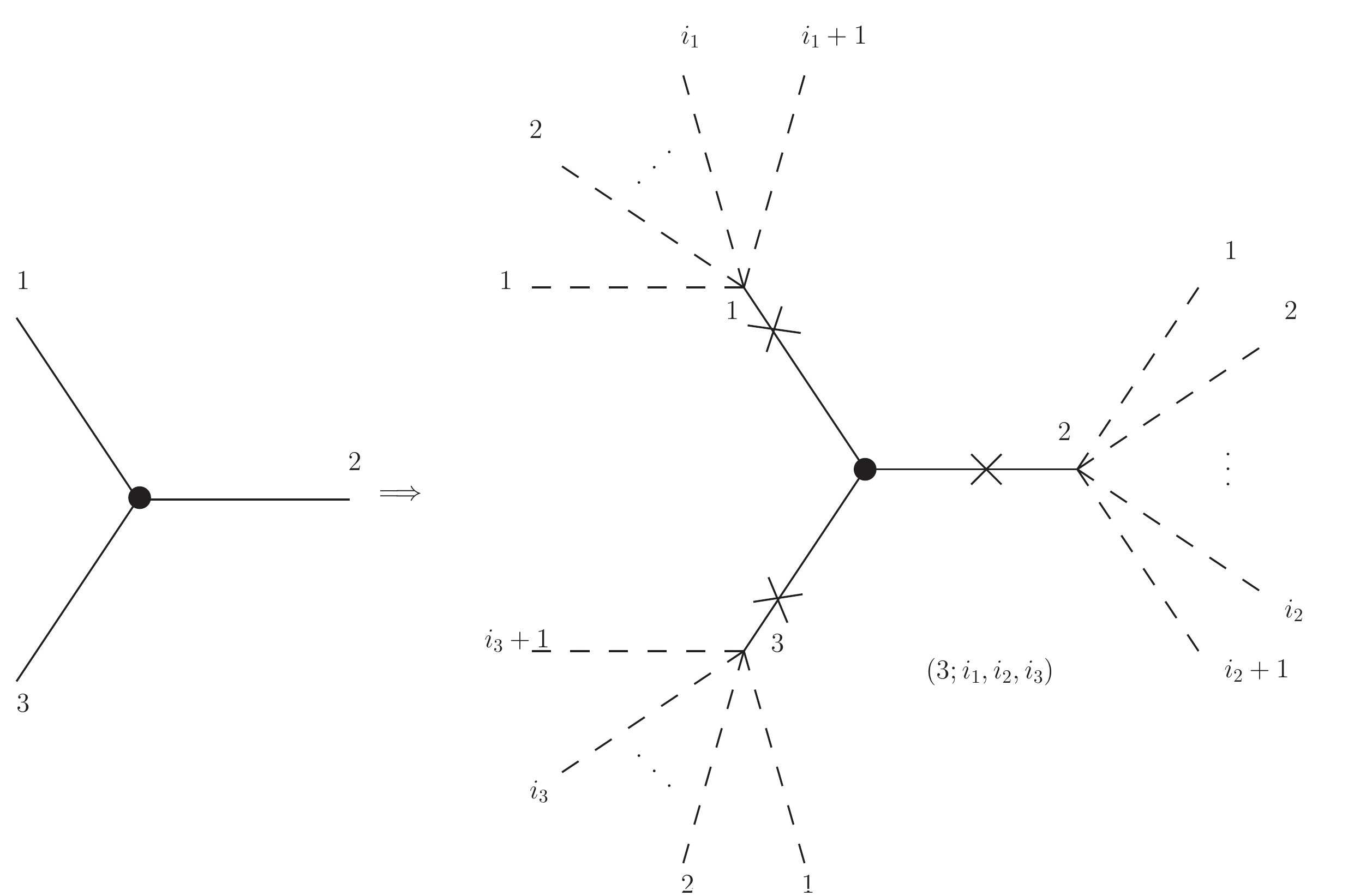}
\end{center}
\caption{The $(3;i_1, i_2, i_3)$ type interaction, with the same notations as in Fig.{\ref{2ij}}.}
\label{3i1i2i3}
\end{figure*}

\begin{figure*}
\begin{center}
\includegraphics[width=8cm,angle=0]{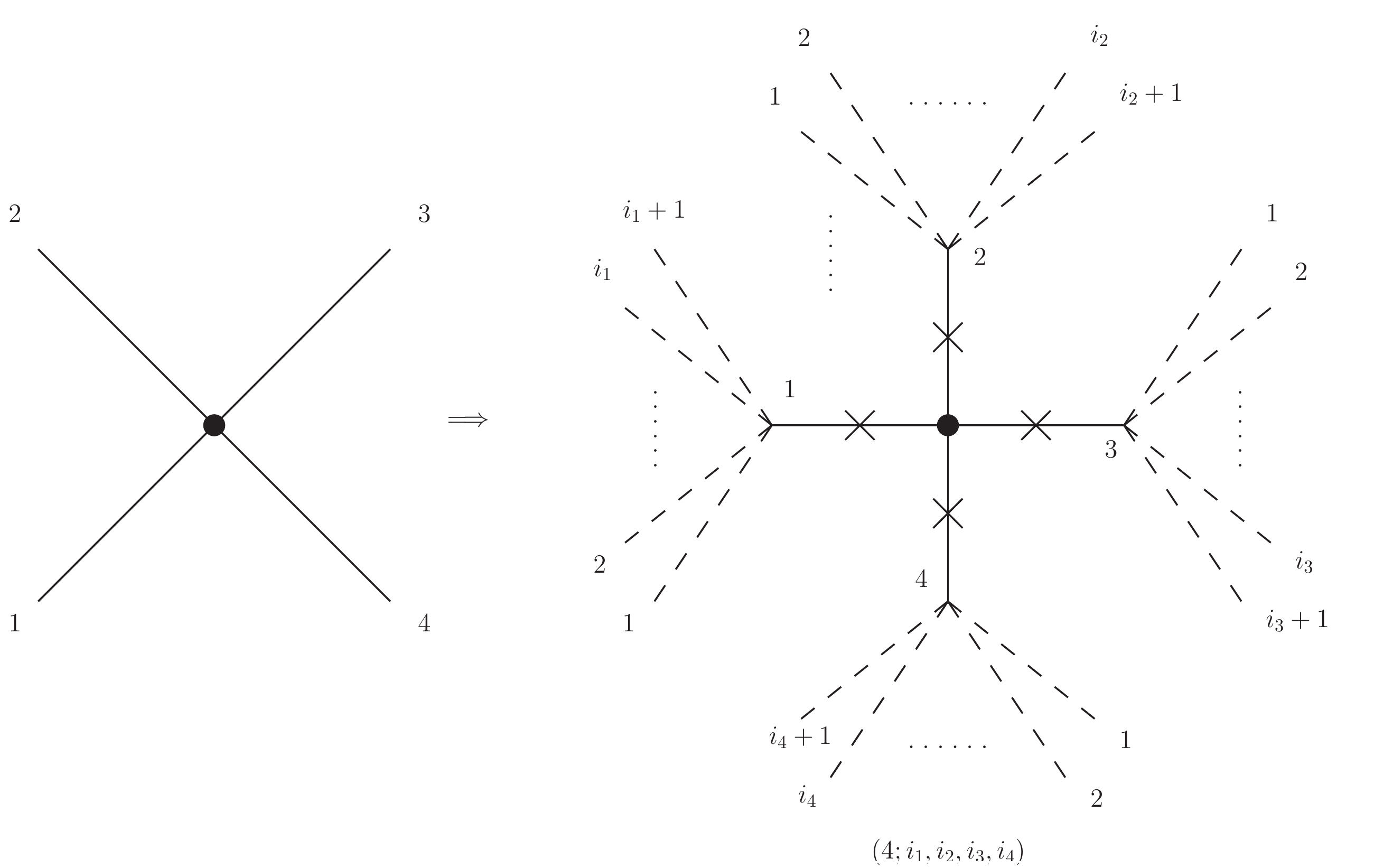}
\end{center}
\caption{The $(4;i_1, i_2, i_3,i_4)$ type interaction, with the same notations as in Fig.{\ref{2ij}}.}
\label{4i1i2i3i4}
\end{figure*}

Relevant terms for each $n$ can then be categorized into three types according to their origins in the SW unmapped theory:

\begin{enumerate}
\item Those by attaching the $i$th order SW map and the $j$th order SW map to a quadratic structure and having $i+j=n-2$ (Fig.~\ref{2ij});

\item those by lifting the three-photon vertex with the 3-SW maps satisfying $i_1+i_2+i_3=n-3$ (Fig.~\ref{3i1i2i3});

\item and those by lifting the four-photon vertex with the 4-SW maps satisfying $i_1+i_2+i_3+i_4=n-4$ (Fig.~\ref{4i1i2i3i4}).

\end{enumerate}
We may denote the first type as $(2; i, j)$, second $(3;i_1, i_2, i_3)$ and third as $(4;i_1, i_2, i_3, i_4)$. 
We also require $j>i$ in $(2; i, j)$, which can be achieved via integration by part without ambiguity thanking to 
the quadratic backbone. We introduce a set of fully ordered Feynman diagrams equivalent/schematic diagrams as diagramatic illustration of these vertex terms. The full Feynman diagram vertices can be expressed by summing over all external leg permuations of all relevant schematic diagrams.

Next, let us consider the  $(2; 0, n-2)$ type:
\begin{equation}
(2; 0, n-2)=(\partial_\mu\partial^\mu a^\nu-\partial_\mu\partial^\nu a^\mu)\cdot A^{(n-2)}_\nu.
\label{3}
\end{equation}
We notice that this term bears the form of a free field equation multiplying an $(n-2)$th-order SW map of 
the NC gauge field $A_\nu$. One can immediately recognize that such vertex does not contribute to 
the $n$-photon scattering amplitude because of the field equation. There are still more intriguing consequences of such terms as we are going to see next.

\subsection{The $\gamma\gamma\to\gamma\gamma$ scattering amplitude}

\begin{figure*}[t]
\begin{center}
\includegraphics[width=14cm,angle=0]{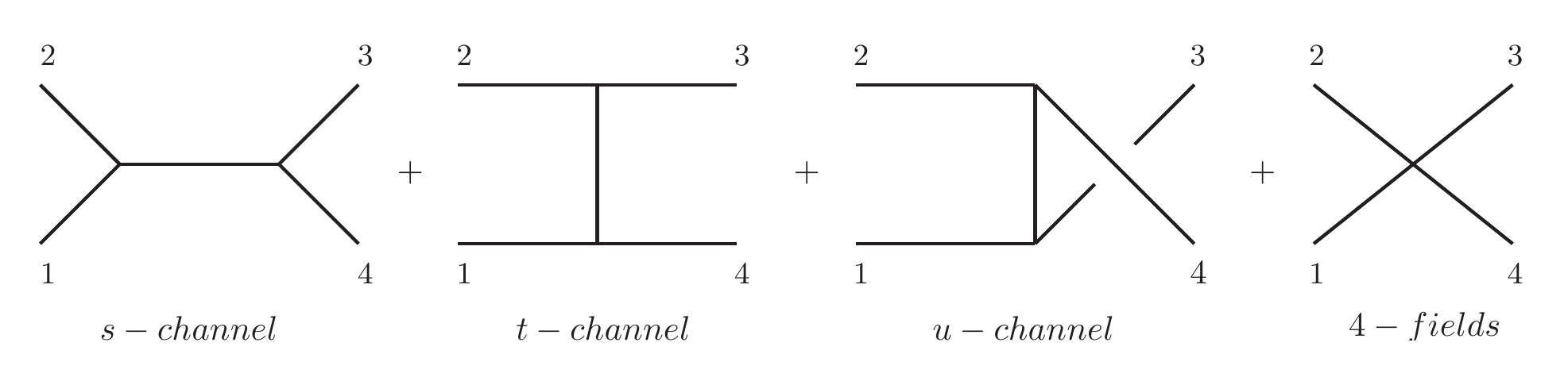}
\end{center}
\caption{Generic diagram contributions, containing the three- and four-field vertices, to the two-by-two $(2\to 2$) scatterings. Full lines denote arbitrary quantum fields.}
\label{Figs-t-u-4SWInvariance}
\end{figure*}

\begin{figure*}
\begin{center}
\includegraphics[width=10cm,angle=0]{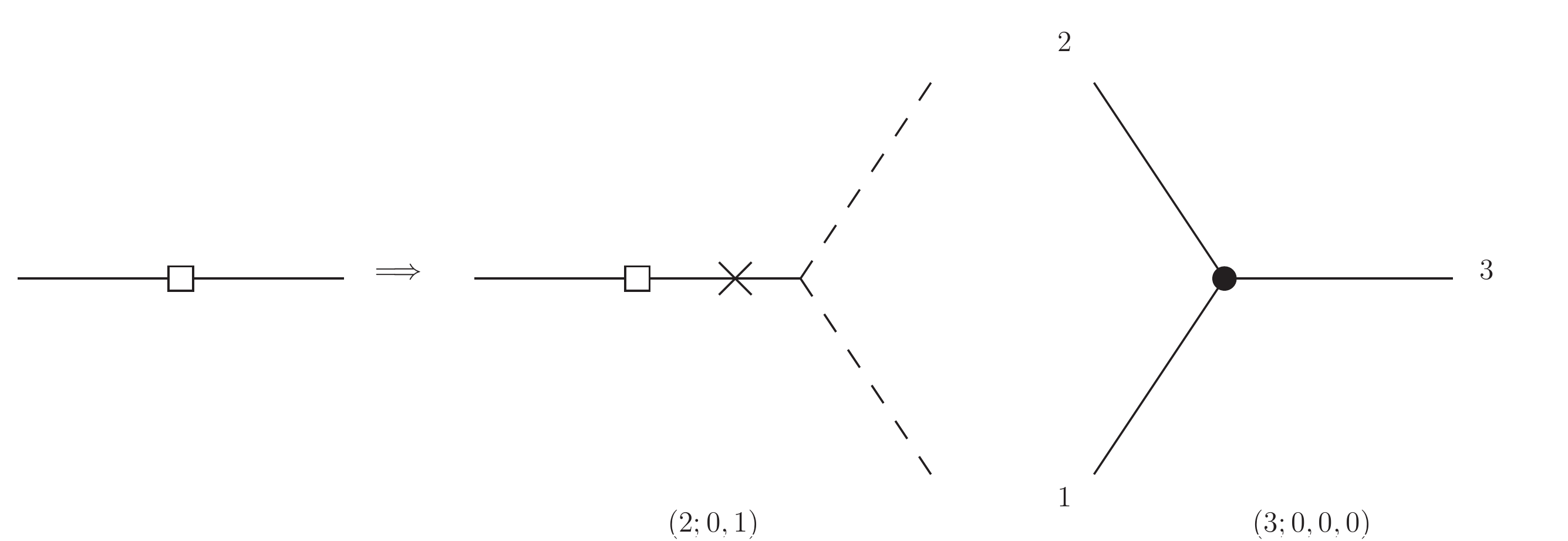}
\end{center}
\caption{The three-photon schematic vertices in SW mapped NCQED}
\label{3photon}
\end{figure*}

\begin{figure*}
\begin{center}
\includegraphics[width=10cm,angle=0]{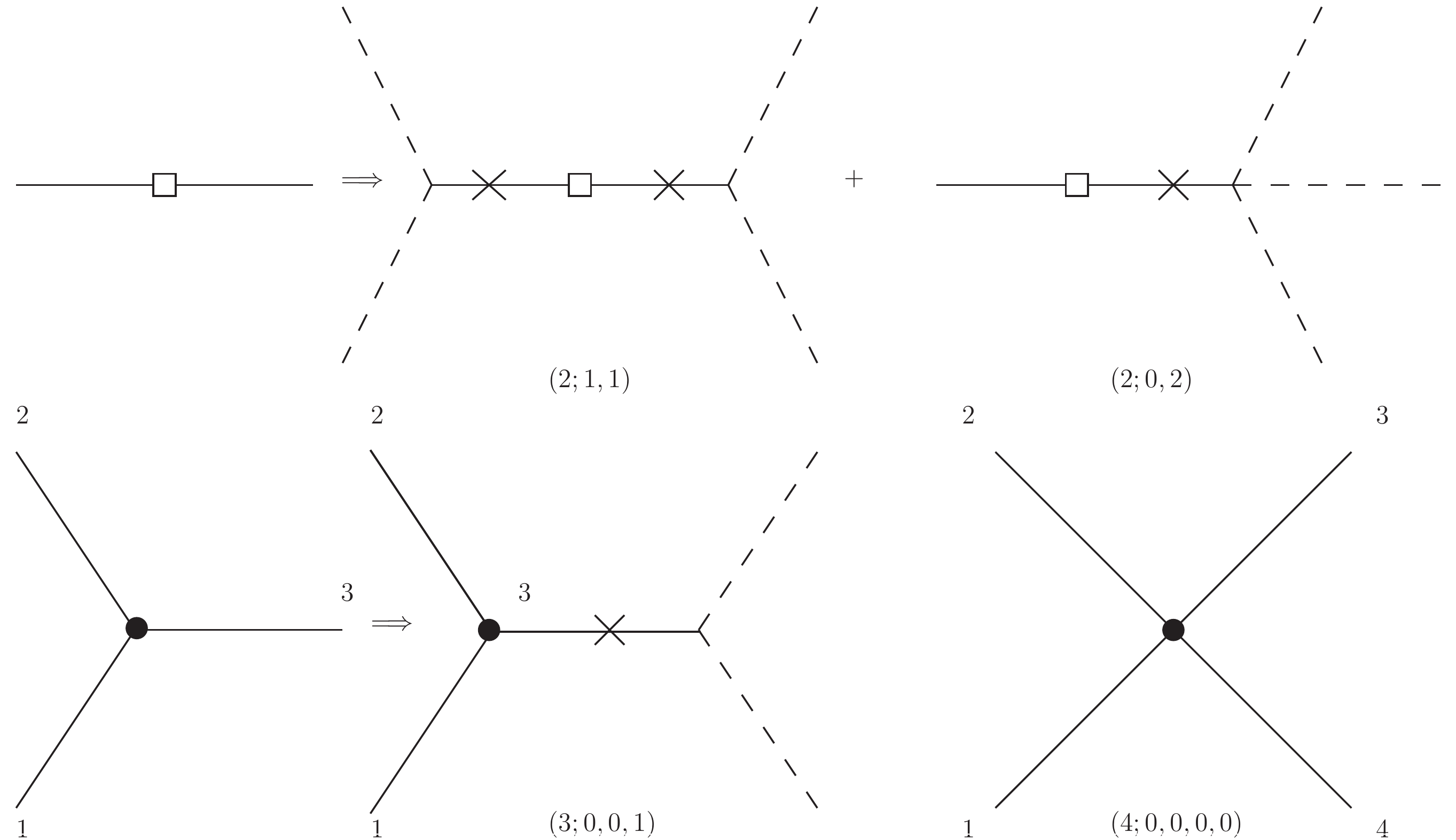}
\end{center}
\caption{The four-photon schematic vertices in SW mapped NCQED}
\label{4photon}
\end{figure*}

As starting example, let us reexamine the SW map invariance of $\gamma\gamma\to\gamma\gamma$ scattering amplitude found in \cite{Latas:2020nji}. Only the three- and four-photon vertices -- ($^1V_{a^3}$) and ($^1V_{a^4}$) -- from minimal/first model defined in Eq.(\ref{NCminAction}) are relevant to this process described in Fig.~\ref{Figs-t-u-4SWInvariance}. 
They can be expressed by the following terms
\begin{gather}
^1V_{a^3}\equiv(2;0,1)+(3;0,0,0),
\label{4}\\
^1V_{a^4}
\equiv(2;0,2)+(2;1,1)+(3;1,0,0)+(3;0,1,0)+(3;0,0,1)+(4;0,0,0,0),
\label{5}
\end{gather}
and be illustrated by the schematic diagrams listed in Figs.~\ref{3photon} and~\ref{4photon}. Notice that we have fixed the order of photon lines completely and neglected all the permutations; they can be reintroduced later without affecting the cancellation of SW map contributions we would like to prove.

\begin{figure*}
\begin{center}
\includegraphics[width=10cm,angle=0]{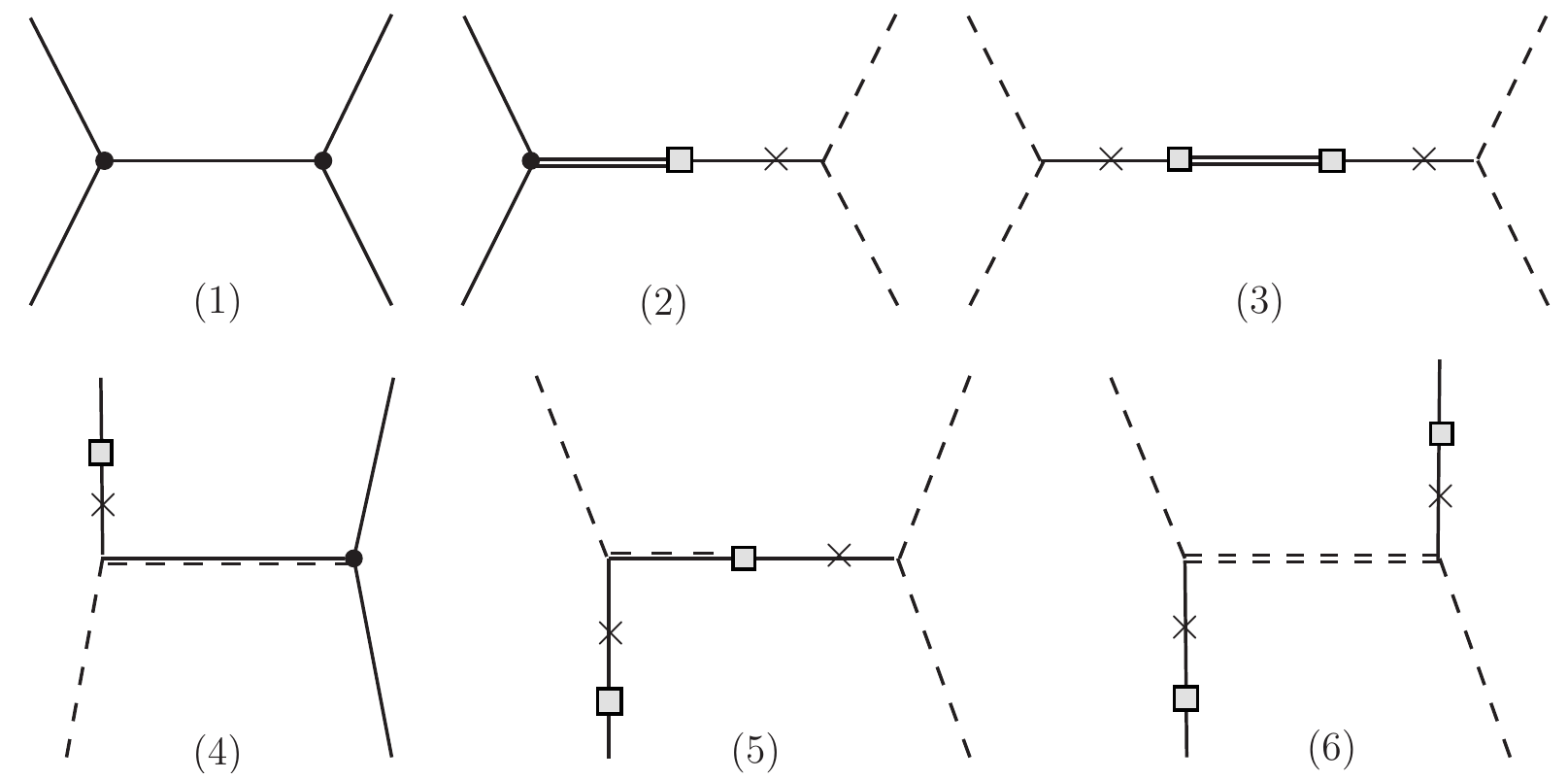}
\end{center}
\caption{The four-photon schematic diagrams which we call the nonvertex type. Solid double lines are used to mark 
the zero-zero type propagator connections here.}
\label{nonvertex}
\end{figure*}

We then consider the Feynman diagrams for scattering amplitudes and expand them into schematic diagrams. These include the four-point vertices in Fig.~\ref{4photon} and nonvertex diagrams in Fig.~\ref{nonvertex}. 

\begin{figure*}
\begin{center}
\includegraphics[width=10cm,angle=0]{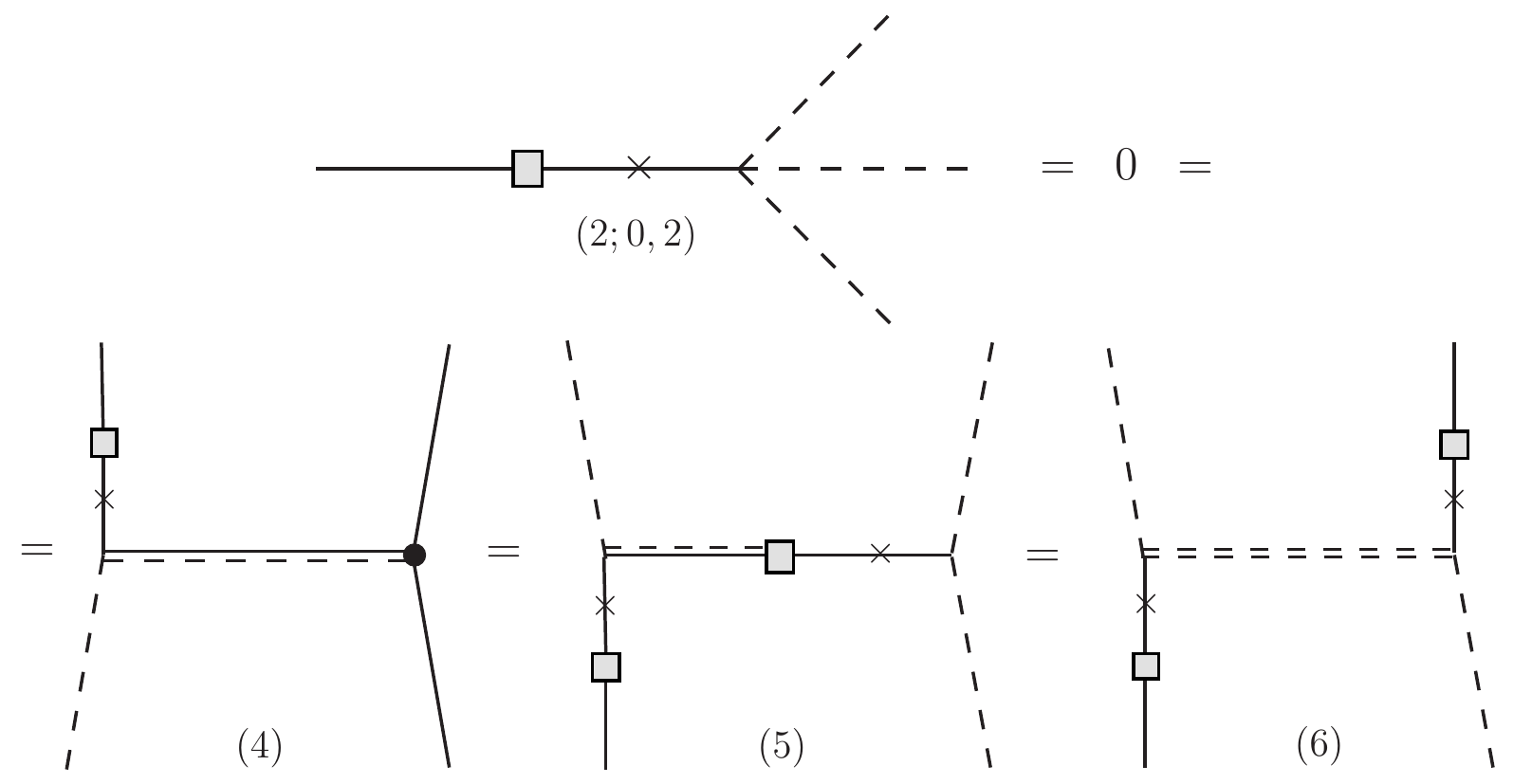}
\end{center}
\caption{Vanishing parts of the SW map induced contributions to  the $2\to2$
($\gamma\gamma\to\gamma\gamma$) scattering amplitudes in the SW mapped NCQED. }
\label{4photoncancellation1}
\end{figure*}

We can first eliminate $(2;0,2)$ as it vanishes on shell. 
Next we realize that the vertex $(2;0,1)$ must have the leg from zeroth order SW map, i.e. with a field equation attached, connected to an internal line in any term/diagram including this vertex, otherwise the resulting term vanishes on shell too. The schematic diagrams which are vanishing on shell because of the field equation in $(2;0,1)$ are illustrated in Fig.~\ref{4photoncancellation1}. Once $(2;0,1)$ is connected to another (full) vertex $V$ via the the leg from the zeroth- order SW map, the the second term in the right-hand parenthesis of \eqref{3} does not contribute because of the property~\eqref{property}, i.e.,
\begin{equation}
(2; 0, 1)\equiv\partial_\mu\partial^\mu a^\nu\cdot A^{(1)}_\nu.
\label{simplified}
\end{equation}

\begin{figure*}
\begin{center}
\includegraphics[width=10cm,angle=0]{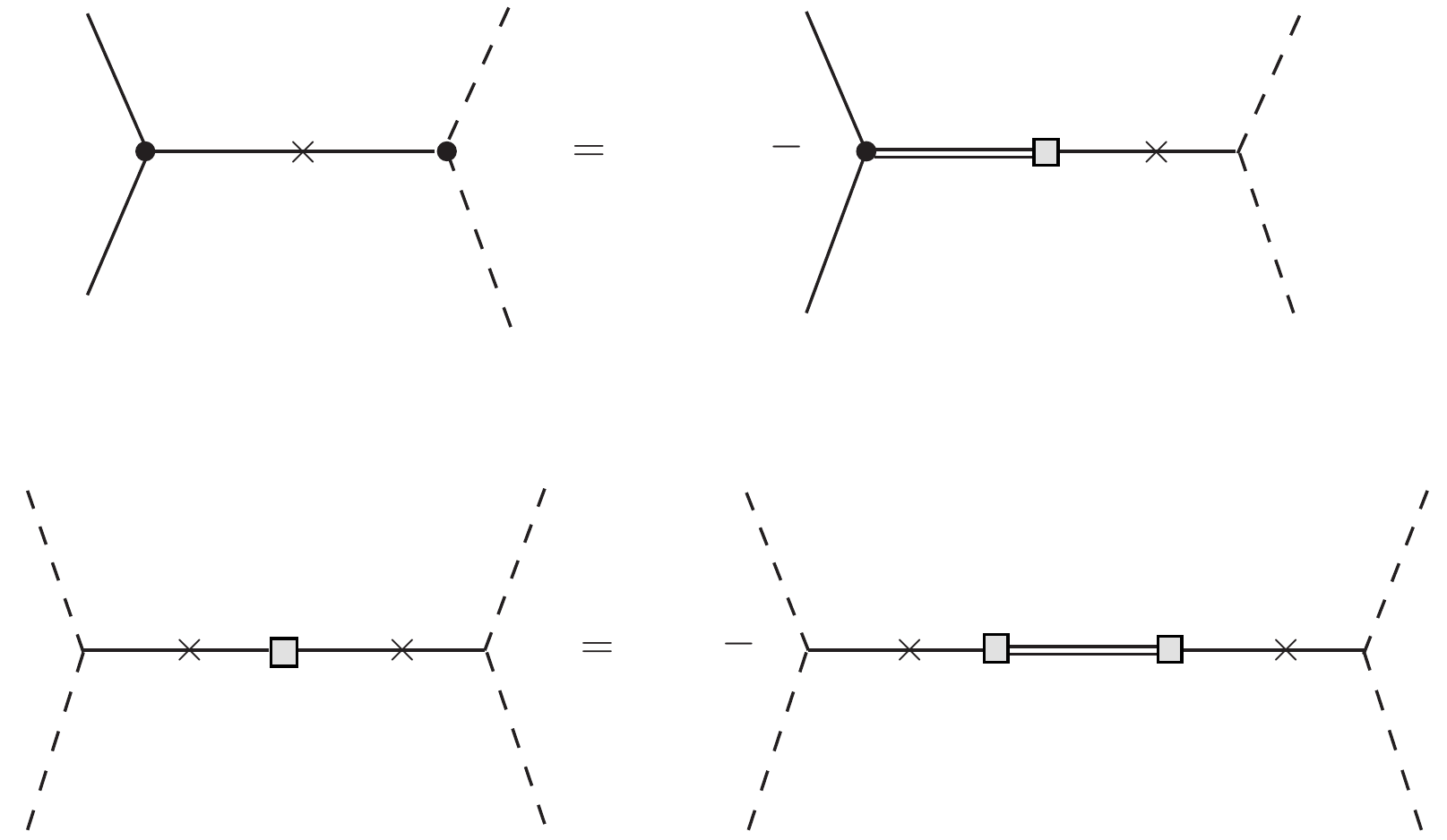}
\end{center}
\caption{Cancellation of the SW map induced contributions to  $2\to2$ 
scattering amplitude  in SW mapped NCQED.  
}
\label{4photoncancellation2}
\end{figure*}

Now we consider attaching the simplified $(2; 0, 1)$, i.e.~\eqref{simplified}, to another vertex for four photon scattering. 
The zeroth leg, i.e., the field equation part must be attached to the internal propagator. There are two possibilities: 
One is to have two $(2;0,1)$ terms connected together via their zeroth legs, the other is to connect it to a $(3;0,0,0)$ term. In both cases, the zeroth leg is connected to another zeroth leg. We mark such connections by a contraction line between terms, and by a double solid line in the schematic diagram. We then notice the following important relations, 
\begin{gather}
\acontraction{(2;}{0}{,1)(2;}{0}(2;0,1)(2;0,1)= -(2;1,1),
\label{cancellation1}
\\
\acontraction{(2;}{0}{,1)(3;0,0,}{0}(2;0,1)(3;0,0,0)= -(3;0,0,1),
\label{cancellation2}
\end{gather} 
given in the language of Wick contractions. This enables the cancellation of all SW map induced contributions, 
whose procedures are given diagrammatically in the 
Fig.~\ref{4photoncancellation2}.

It is also plain that $(3;1,0,0)$ and $(3;0,1,0)$ are canceled by $\acontraction{(2;}{0}{,1)(3;}{0}(2;0,1)(3;0,0,0)$ 
and $\acontraction{(2;}{0}{,1)(3;0,}{0}(2;0,1)(3;0,0,0)$, respectively. These cancellations then leave the diagrams which 
are not induced by SW map the only nonvanishing contribution to four-photon scattering amplitude.

\subsection{Generalization to $n$-photon scattering amplitudes}

To generalize our cancellation results to all tree-level photon scattering amplitudes, we first notice that all $(2;i>0,j>i>0)$, $(3;i_1,i_2,i_3)$ with $i_1+i_2+i_3>0$, and $(4;i_1,i_2,i_3,i_4)$ with $i_1+i_2+i_3+i_4>0$ vertices can be canceled by summing with schematic diagrams which have crossed legs replaced by double-line-box-cross propagators one by one until all of them. These relations can be written in terms of zero-zero contractions as follows: First, we define the vanishing sum
\begin{equation}
0\equiv[2;i>0,j>i>0]=(2;i>0,j>i>0)+\acontraction{(2;}{0}{,i)(2;}{0}(2;0,i)(2;0,j),
\label{9}
\end{equation}
then
\begin{equation}
\begin{split}
0\equiv&[3;i_1,i_2,i_3]=(3;i_1,i_2,i_3)
+\acontraction{(2;}{0}{,i_1)(3;}{0}(2;0,i_1)(3;0,i_2,i_3)
\\&+\acontraction{(2;}{0}{,i_2)(3;i_1,}{0}(2;0,i_2)(3;i_1,0,i_3)
+\acontraction{(2;}{0}{,i_3)(3;i_1,i_2,}{0}(2;0,i_3)(3;i_1,i_2,0)
\\&+
\acontraction{(2;}{0}{,i_2)(2;0,i_3)(3;i_1,}{0}
\bcontraction{(2;0,i_2)(2;}{0}{,i_2)(3;i_1,0,}{0}
(2;0,i_2)(2;0,i_3)(3;i_1,0,0)
+
\acontraction{(2;}{0}{,i_1)(2;0,i_3)(3;}{0}
\bcontraction{(2;0,i_1)(2;}{0}{,i_3)(3;0,i_2,}{0}
(2;0,i_1)(2;0,i_3)(3;0,i_2,0)
\\&+
\acontraction{(2;}{0}{,i_1)(2;0,i_2)(3;}{0}
\bcontraction{(2;0,i_1)(2;}{0}{,i_2)(3;0,}{0}
(2;0,i_1)(2;0,i_2)(3;0,0,i_3)
+
\acontraction{(2;}{0}{,i_1)(2;0,i_2)(2;0,i_3)(3;}{0}
\acontraction[2ex]{(2;0,i_1)(2;}{0}{,i_2)(2;0,i_3)(3;0,}{0}
\bcontraction{(2;0,i_1)(2;0,i_2)(2;}{0}{,i_3)(3;0,0,}{0}
(2;0,i_1)(2;0,i_2)(2;0,i_3)(3;0,0,0),
\end{split}
\label{10}
\end{equation}
and finally 
\begin{equation}
\begin{split}
0\equiv&[4;i_1,i_2,i_3,i_4]=(4;i_1,i_2,i_3,i_4)
+
\acontraction{(2;}{0}{,i_1)(4;}{0}
(2;0,i_1)(4;0,i_2,i_3,i_4)
\\&+
\acontraction{(2;}{0}{,i_2)(4;i_1,}{0}
(2;0,i_2)(4;i_1,0,i_3,i_4)
+
\acontraction{(2;}{0}{,i_3)(4;i_1,i_2,}{0}
(2;0,i_3)(4;i_1,i_2,0,i_4)
+
\acontraction{(2;}{0}{,i_4)(4;i_1,i_2,i_3,}{0}
(2;0,i_4)(4;i_1,i_2,i_3,0)
\\&+
\acontraction{(2;}{0}{,i_1)(2;0,i_2)(4;}{0}
\bcontraction{(2;0,i_1)(2;}{0}{,i_2)(4;0,}{0}
(2;0,i_1)(2;0,i_2)(4;0,0,i_3,i_4)
+
\acontraction{(2;}{0}{,i_1)(2;0,i_3)(4;}{0}
\bcontraction{(2;0,i_1)(2;}{0}{,i_3)(4;0,i_2,}{0}
(2;0,i_1)(2;0,i_3)(4;0,i_2,0,i_4)
\\&+
\acontraction{(2;}{0}{,i_1)(2;0,i_4)(4;}{0}
\bcontraction{(2;0,i_1)(2;}{0}{,i_4)(4;0,i_2,i_3,}{0}
(2;0,i_1)(2;0,i_4)(4;0,i_2,i_3,0)
+
\acontraction{(2;}{0}{,i_2)(2;0,i_3)(4;i_1,}{0}
\bcontraction{(2;0,i_2)(2;}{0}{,i_3)(4;i_1,0,}{0}
(2;0,i_2)(2;0,i_3)(4;i_1,0,0,i_4)
\\&+
\acontraction{(2;}{0}{,i_2)(2;0,i_4)(4;i_1,}{0}
\bcontraction{(2;0,i_2)(2;}{0}{,i_4)(4;i_1,0,i_3,}{0}
(2;0,i_2)(2;0,i_4)(4;i_1,0,i_3,0)
+
\acontraction{(2;}{0}{,i_3)(2;0,i_4)(4;i_1,i_2,}{0}
\bcontraction{(2;0,i_3)(2;}{0}{,i_4)(4;i_1,i_2,0,}{0}
(2;0,i_3)(2;0,i_4)(4;i_1,i_2,0,0)
\\&+
\acontraction{(2;}{0}{,i_1)(2;0,i_2)(2;0,i_3)(3;}{0}
\acontraction[2ex]{(2;0,i_1)(2;}{0}{,i_2)(2;0,i_3)(4;0,}{0}
\bcontraction{(2;0,i_1)(2;0,i_2)(2;}{0}{,i_3)(4;0,0,}{0}
(2;0,i_1)(2;0,i_2)(2;0,i_3)(4;0,0,0,i_4)
+
\acontraction{(2;}{0}{,i_1)(2;0,i_2)(2;0,i_4)(4;}{0}
\acontraction[2ex]{(2;0,i_1)(2;}{0}{,i_2)(2;0,i_4)(4;0,}{0}
\bcontraction{(2;0,i_1)(2;0,i_2)(2;}{0}{,i_4)(4;0,0,i_3,}{0}
(2;0,i_1)(2;0,i_2)(2;0,i_4)(4;0,0,i_3,0)
\\&+
\acontraction{(2;}{0}{,i_1)(2;0,i_3)(2;0,i_4)(4;}{0}
\acontraction[2ex]{(2;0,i_1)(2;}{0}{,i_3)(2;0,i_4)(4;0,i_2,}{0}
\bcontraction{(2;0,i_1)(2;0,i_3)(2;}{0}{,i_4)(4;0,i_2,0,}{0}
(2;0,i_1)(2;0,i_3)(2;0,i_4)(4;0,i_2,0,0)
+
\acontraction{(2;}{0}{,i_2)(2;0,i_3)(2;0,i_4)(4;i_1,}{0}
\acontraction[2ex]{(2;0,i_2)(2;}{0}{,i_3)(2;0,i_4)(4;i_1,0,}{0}
\bcontraction{(2;0,i_2)(2;0,i_3)(2;}{0}{,i_4)(4;i_1,0,0,}{0}
(2;0,i_2)(2;0,i_3)(2;0,i_4)(4;i_1,0,0,0)
\\&+
\acontraction{(2;}{0}{,i_1)(2;0,i_2)(2;0,i_3)(2;0,i_4)(4;}{0}
\acontraction[2ex]{(2;0,i_1)(2;}{0}{,i_2)(2;0,i_3)(2;0,i_4)(4;0,}{0}
\bcontraction{(2;0,i_1)(2;0,i_2)(2;}{0}{,i_3)(2;0,i_4)(4;0,0,}{0}
\bcontraction[2ex]{(2;0,i_1)(2;0,i_2)(2;0,i_3)(2;}{0}{,i_4)(4;0,0,0,}{0}
(2;0,i_1)(2;0,i_2)(2;0,i_3)(2;0,i_4)(4;0,0,0,0).
\end{split}
\label{11}
\end{equation}
Above we have introduced square brackets $[2;i>0,j>i>0]$, $[3;i_1,i_2,i_3]$ and $[4;i_1,i_2,i_3,i_4]$ to denote these vanishing sums. We realize that we can treat these vanishing sums as effective vertices. All double-solid internal lines are then absorbed into these effective vertices. Introducing also the compatible notation
\begin{equation}
[2;0,i]\equiv(2;0,i),\quad [3;0,0,0]\equiv(3;0,0,0),\quad [4;0,0,0,0]\equiv(4;0,0,0,0),
\label{12}
\end{equation}
we can define a modified type of schematic diagrams by demanding that the new, box-bracket, vertices can only 
be connected by internal propagators which are not the double solid line type. Usual $n$-photon correlation 
functions can still be expressed in terms of these new schematic diagrams since all terms in the definition 
of box-bracket vertices are equivalent to each other if we restrict the internal propagator connections as stated.

It is then straightforward to see that one new schematic diagram vanishes if it contains any of the $[2;i>0,j>i>0]$, 
$[3;i_1,i_2,i_3]$ with $i_1+i_2+i_3>0$, and $[4;i_1,i_2,i_3,i_4]$ with $i_1+i_2+i_3+i_4>0$ vertices. 
So the nonvanishing new schematic diagrams can only be built up by $[2;0,i]$, $ [3;0,0,0]$, and $[4;0,0,0,0]$ vertices 
and internal propagators which are not the double solid line type. Now assume that one schematic diagram is from the SW map. Then it must contain at least one $[2;0,i]$ vertex. The leg of this $[2;0,i]$ vertex with a field equation attached must then not be external, as otherwise that diagram vanishes on shell. This leg can not connect to a $ [3;0,0,0]$ or $[4;0,0,0,0]$ vertex either because of the modified Feynman rule we just imposed. Thus the only choice is to connect this leg with one dashed line of another $[2;0,j]$ vertex. Then the next vertex's zeroth leg must find one more dashed line of one more vertex to connect. Now this chain of vertices must end somewhere for any scattering amplitude with a finite number of photons. Then the zeroth leg of the last vertex on this chain has to stay external, which forces the whole diagram to vanish onshell. Therefore all SW map-induced contributions must vanish on shell in the very end.

\subsection{Scattering amplitudes including fermions}

To extend the above procedure to fermions it is necessary to use a general result in QED, that is: ``Eq. \eqref{property} type of off shell contraction vanish for a general QED amplitude if and only if all charged particles are on shell''
(Weinberg, Vol. 1, \S~10.5 \cite{Weinberg}). SW map-generated vertices can now come from the quadratic $\bar\Psi\Psi$ and cubic $\bar\Psi\slashed A\Psi$ backbone terms. We can introduce the notation $(\bar\Psi\Psi;\bar i, i)$ and $(\bar\Psi\slashed A\Psi;\bar i, j, i)$ to denote the SW mapped terms. Here in the first term $\bar i$ denotes the order of SW map for $\bar\Psi$ and $i$ for $\Psi$, while in the second $\bar i$ for $\bar\Psi$, $j$ for $A^\mu$ and $i$ for $\Psi$, respectively.

Vanishing effective vertices $[\bar\Psi\Psi;\bar i, i]$ and $[\bar\Psi\slashed A\Psi;\bar i, j, i]$ can then 
be built using $(\bar\Psi\Psi;\bar 0, i)$, $(\bar\Psi\Psi;\bar i, 0)$, and $(2;0, i)$ elements. We have
\begin{equation}
0\equiv[\bar\Psi\Psi;\bar i, i]=(\bar\Psi\Psi;\bar i, i)+\acontraction{(\bar\Psi\Psi;\bar{i},}{0}{)(\bar\Psi\Psi;}{\bar{0}}(\bar\Psi\Psi;\bar{i},0)(\bar\Psi\Psi;\bar{0},i), 
\label{13}
\end{equation}
and
\begin{equation}
\begin{split}
0\equiv&[\bar\Psi\slashed A\Psi;\bar i, j, i]=(\bar\Psi\slashed A\Psi;\bar i, j, i)+\acontraction{(\bar\Psi\Psi;\bar{i},}{0}{)(\bar\Psi\slashed A\Psi;}{\bar 0}(\bar\Psi\Psi;\bar{i},0)(\bar\Psi\slashed A\Psi;\bar 0, j, i)
+\bcontraction{(2;}{0}{,j)(\bar\Psi\slashed A\Psi;\bar i,}{0}(2;0,j)(\bar\Psi\slashed A\Psi;\bar i, 0, i)
\\&+\acontraction{(\bar\Psi\slashed A\Psi;\bar i, j,}{0}{)(\bar\Psi\Psi;}{\bar{0}}(\bar\Psi\slashed A\Psi;\bar i, j, 0)(\bar\Psi\Psi;\bar{0},i)
+
\acontraction{(\bar\Psi\Psi;\bar{i},}{0}{)(2;0,j)(\bar\Psi\slashed A\Psi;}{\bar 0}
\bcontraction{(\bar\Psi\Psi;\bar{i},0)(2;}{0}{,j)(\bar\Psi\slashed A\Psi;\bar 0,}{0}
(\bar\Psi\Psi;\bar{i},0)(2;0,j)(\bar\Psi\slashed A\Psi;\bar 0, 0, i)
\\&+
\bcontraction{(2;}{0}{,j)(\bar\Psi\slashed A\Psi;\bar i,}{0}
\acontraction{(2;0,j)(\bar\Psi\slashed A\Psi;\bar i, 0,}{0}{)(\bar\Psi\Psi;}{\bar{0}}
(2;0,j)(\bar\Psi\slashed A\Psi;\bar i, 0, 0)(\bar\Psi\Psi;\bar{0},i)
+
\acontraction{(\bar\Psi\Psi;\bar{i},}{0}{)(\bar\Psi\slashed A\Psi;}{\bar 0}
\acontraction{(\bar\Psi\Psi;\bar{i},0)(\bar\Psi\slashed A\Psi;\bar 0, j,}{0}{)(\bar\Psi\Psi;}{\bar{0}}
(\bar\Psi\Psi;\bar{i},0)(\bar\Psi\slashed A\Psi;\bar 0, j, 0)(\bar\Psi\Psi;\bar{0},i)
\\&+
\acontraction{(\bar\Psi\Psi;\bar{i},}{0}{)(2;0,j)(\bar\Psi\slashed A\Psi;}{\bar 0}
\bcontraction{(\bar\Psi\Psi;\bar{i},0)(2;}{0}{,j)(\bar\Psi\slashed A\Psi;\bar 0,}{0}
\acontraction{(\bar\Psi\Psi;\bar{i},0)(2;0,j)(\bar\Psi\slashed A\Psi;\bar 0, 0,}{0}{)(\bar\Psi\Psi;}{\bar{0}}
(\bar\Psi\Psi;\bar{i},0)(2;0,j)(\bar\Psi\slashed A\Psi;\bar 0, 0, 0)(\bar\Psi\Psi;\bar{0},i).
\end{split}
\label{14}
\end{equation}
Remaining arguments then follow the photon only case, except that we need all external fermions to be on shell.

\section{Moyal deformed QED built by irreversible SW map: second NCQED model}

In previous sections we have proven that all tree-level scattering amplitudes of the Moyal U(1) NCQED model ~\eqref{NCminAction} satisfy the Seiberg-Witten map invariance, i.e. that all tree-level scattering amplitudes are the same before and after performing SW map. On the other hand, it is long known that Moyal deformed NCQED can bear a more general definition than minimal/first model~\eqref{NCminAction}, if irreversible SW map is used. This more general definition allows commutative matter fields with different charges to couple to the same commutative U(1) gauge field by building multiple NC gauge fields required from the same commutative gauge field~\cite{Horvat:2011qn}. Such constructions were relevant for proper NC deformations of commutative theories with comprhenstive U(1) charge assignments~\cite{Horvat:2011qn,Calmet:2001na,Aschieri:2002mc,Martin:2010ng}. This procedure yields a commutative U(1) gauge invariant classical theory, yet breaks the formal equivalence relation between the theories before and after SW map~\cite{Martin:2016hji,Martin:2016saw}, as they have different degrees of freedom now. In this section we discuss how the tree-level scattering amplitudes in such theories can, in turn, differ from the Moyal NCQED without SW map and/or with reversible SW map.

We consider a second model built by irreversible $\theta$-exact SW map, in order to compare with the minimal model \eqref{NCminAction}.
This model has a fermion sector with two NC fermion fields $\Psi_1$ and $\Psi_2$, with charges $e$ and $\kappa e$ respectively, for simplicity \cite{Horvat:2011qn,Horvat:2012vn}. The gauge sector contains, in turn, two terms generated by two different NC gauge fields. The formal NC action of such a model can be constructed as follows:
\begin{equation}
\begin{split}
&^2S\big[\Psi_1(\psi_1, e a_\mu),\Psi_2(\psi_2, \kappa e a_\mu),{A_1}_\mu(e a_\mu),{A_2}_\mu(\kappa e a_\mu)\big]
\\&=\sum\limits_{i=1}^2\int\bar\Psi_i \star(i\slashed{D}^{(i)}-m)\Psi_i 
-\frac{1}{4G^2}\int \Big(F_{\mu\nu}[{A_1}_\mu(e a_\mu)]\star F^{\mu\nu}[{A_1}_\mu(e a_\mu)]
+\lambda F_{\mu\nu}[{A_2}_\mu(\kappa e a_\mu)]\star F^{\mu\nu}[{A_2}_\mu(\kappa e a_\mu)]
\Big),
\label{SWAction}
\end{split}
\end{equation}
where
\begin{equation}
D^{(i)}_\mu\Psi_i=\partial_\mu\Psi_i-i{A_i}_\mu\star\Psi_i.
\label{16}
\end{equation}
The normalization constant $G^2$ is set to be
\begin{equation}
G^2=e^2(1+\lambda\kappa^2),
\label{17}
\end{equation}
which ensures that {($\ref{SWAction}$)} has the regular commutative limit when $\theta\to 0$.
And the weight factor $\lambda$ between two different parts of the gauge sector is free. One can then notice that the deviation from the minimal model \eqref{NCminAction} comes from the fact that one can only build up vertices based purely on ${A_1}_\mu$ or ${A_2}_\mu$. On the other hand, diagrams can freely connect vertices from both ${A_i}_\mu$ parts, since in the end they are expressed in terms of the same commutative gauge field  $a_\mu$. We then expect modification to the fermion-photon scattering and photon-photon scattering amplitudes of the two-by-two scattering processes.

Since the photon interaction vertices are now linear combination of two separated parts, we can generalize the notation in previous sections by marking the origin of each term. We can write the three- and four-field vertices from the second model (\ref{SWAction}) as follows:
\begin{gather}
^2V_{a^3}=(A_1|2;0,1)+(A_1|3;0,0,0)+(A_2|2;0,1)+(A_2|3;0,0,0),
\label{18}\\
\begin{split}
^2V_{a^4}=&(A_1|2;0,2)+(A_1|2;1,1)+(A_1|3;0,0,1)+(A_1|4;0,0,0,0)
\\&+(A_2|2;0,2)+(A_2|2;1,1)+(A_2|3;0,0,1)+(A_2|4;0,0,0,0),
\end{split}
\label{19}
\end{gather}
and
\begin{gather}
^2V_{\bar\psi_i a\psi_i}=(\bar\Psi_i\Psi_i;\bar 0, 1)+(\bar\Psi_i\Psi_i;\bar 1, 0)+(\bar\Psi_i\slashed A_i\Psi_i;\bar 0, 0, 0),
\label{20}\\
\begin{split}
^2V_{\bar\psi_i a^2\psi_i}=&(\bar\Psi_i\Psi_i;\bar 0, 2)+(\bar\Psi_i\Psi_i;\bar 2, 0)+(\bar\Psi_i\Psi_i;\bar 1, 1)
\\&+(\bar\Psi_i\slashed A_i\Psi_i;\bar 1, 0, 0)+(\bar\Psi_i\slashed A_i\Psi_i;\bar 0, 1, 0)+(\bar\Psi_i\slashed A_i\Psi_i;\bar 0, 0, 1).
\end{split}
\label{21}
\end{gather}
It is not difficult to notice that the SW map used for the second model is the same as for the first model, except for different charge prefactors in front of $a_\mu$ when defining ${A_1}_\mu(e a_\mu)$ and ${A_2}_\mu(\kappa e a_\mu)$. For this reason, each parenthesis term in the vertices \eqref{18} to {\eqref{21}} of the second model is equal to its counterpart in the first model up to a prefactor. These relations can be expressed explicitly for the gauge sector as follows: 
\begin{gather}
(A_2|2;i,j)=\lambda\kappa^{i+j}(A_1|2;i,j)=\frac{\lambda\kappa^{i+j}}{1+\lambda\kappa^2}(2;i,j),
\label{22}\\
(A_2|3;i,j,k)=\lambda\kappa^{i+j+k}(A_1|3;i,j,k)=\frac{\lambda\kappa^{i+j+k}}{1+\lambda\kappa^2}(3;i,j,k),
\label{23}\\
(A_2|4;i,j,k,l)=\lambda\kappa^{i+j+k+l}(A_1|4;i,j,k,l)=\frac{\lambda\kappa^{i+j+k+l}}{1+\lambda\kappa^2}(4;i,j,k,l),
\label{24}
\end{gather}
where $(2;i,j)$, $(3;i,j,k)$, and $(4;i,j,k,l)$ are vertex terms of the first model. 

Similarly, the matter sector vertices of both models are related as follows:
\begin{gather}
^2V_{\bar\psi_2 a^n\psi_2}=\kappa^n \cdot{^2V_{\bar\psi_1 a^n\psi_1}}=\kappa^n \cdot{^1V_{\bar\psi a^n\psi}},
\label{25}
\end{gather}
where $^1V_{\bar\psi a^n\psi}$ denote the vertex terms with $n$ photon field legs and two fermions legs of the minimal/first model~\eqref{NCminAction}, after the SW map was performed, (with the fermions relabeled).

Since the coefficients are different between parentheses, it is reasonable to expect the cancellation of SW map contributions no longer holds in general in the second model. Indeed, a detailed inspection of two-by-two scattering amplitudes of the  the second model~\eqref{SWAction} resulted in that the $\bar\psi_i\psi_i\to\bar\psi_j\psi_j$ (and as well $\psi_i\psi_i\to\psi_j\psi_j$ ) scattering amplitude is, up to an overall factor, the same as $\bar\psi\psi\to\bar\psi\psi$ (and $\psi\psi\to\psi\psi$) of the first model~\eqref{NCminAction}, while for Compton ($\gamma\psi_i\to\gamma\psi_i$) and light-by-light ($\gamma\gamma\to\gamma\gamma$) scatterings we have new terms  because we linearly shifted the photon self-interactions. With the help of relations \eqref{22} to \eqref{25}, we can express the $\gamma\psi_i\to\gamma\psi_i, \;i=1,2$ and $\gamma\gamma\to\gamma\gamma$ scattering amplitudes ($^2\Gamma$) of the second model~\eqref{SWAction} as their counterparts ($^1\Gamma$) in the first model~\eqref{NCminAction}, in which all SW map contributions vanish, plus the irreversible $\theta$-exact SW map induced corrections $\Delta$,
\begin{eqnarray}
^2\Gamma({\gamma\psi_1\to\gamma\psi_1})&=&^1\Gamma({\gamma\psi\to\gamma\psi})+\Delta({\gamma\psi_1\to\gamma\psi_1}),
\label{26}\\
^2\Gamma({\gamma\psi_2\to\gamma\psi_2})&=&\kappa^2\cdot^1\Gamma({\gamma\psi\to\gamma\psi})+\Delta({\gamma\psi_2\to\gamma\psi_2}),
\label{27}\\
^2\Gamma({\gamma\gamma\to\gamma\gamma})&=&^1\Gamma({\gamma\gamma\to\gamma\gamma})+\Delta({\gamma\gamma\to\gamma\gamma}),
\label{28}
\end{eqnarray}
where
\begin{gather}
\Delta({\gamma\psi_1\to\gamma\psi_1})=-\lambda\kappa\Delta({\gamma\psi_2\to\gamma\psi_2})=\frac{\lambda\kappa^2(1-\kappa)}{1+\lambda\kappa^2}(\bar\Psi\slashed{A}\Psi;\bar 0,1,0)\Big|_{\rm on-shell},
\label{29}
\\
\Delta({\gamma\gamma\to\gamma\gamma})=\frac{\lambda^2\kappa^2(1-\kappa)^2}{(1+\lambda\kappa^2)^2}\Big[(2;1,1)+(3;0,0,1)+(3;0,1,0)+(3;1,0,0)\Big]_{\rm on-shell}.
\label{30}
\end{gather}

As expected, the above irreversible SW map induced corrections vanish when $\kappa$ equals to 1 and 0, since $\kappa=1$ makes ${A_1}_\mu$ equal ${A_2}_\mu$ and $\kappa=0$ forces $\psi_2$ to decouple from the gauge field. The fact that $\Delta({\gamma\psi_2\to\gamma\psi_2})$ does not depend linearly on $\lambda$ implies that it is not possible to remove the irreversible $\theta$-exact SW map corrections by adjusting the weight between two parts of the gauge sector. 

The annihilation $\psi_i\bar\psi_i\to \gamma\gamma$ and pair production 
$\gamma\gamma\to\psi_i\bar\psi_i$ amplitudes can be derived from the Compton amplitude by crossing symmetry, so they carry the same $\kappa$ and $\lambda$ dependences as the $\gamma\psi_i\to\gamma\psi_i$ does.

It is very important to note that $\Delta({\gamma\psi_i\to\gamma\psi_i}), \;i=1,2$ and $\Delta({\gamma\gamma\to\gamma\gamma})$ do not contain infrared divergences since they can be expressed as vertices only, i.e. there are no Green functions generating additional singularities. Therefore the leading IR singularities of the differential cross sections from the second model~\eqref{SWAction} remain the same as that of the minimal/first model~\eqref{NCminAction}.

\section{Discussion and conclusion}

In this work we show in two parts the connection between reversibility of Seiberg-Witten maps and an identical relation between the tree-level scettering amplitudes of the Moyal NCQED with and without a SW map.
The first and main part of the manuscript demonstrates that all tree-level scattering amplitudes of a minimal/first NCQED U(1) model with gauge boson and one left-charge fermion before and after a reversible SW map are equal to each other. 
The second part handles one of the opposite scenarios, in which an irreversible SW map is used to define the second NCQED model with two, differently charged, fermions. We find that the aforementioned identity breaks down for the two-by-two photon-fermion and photon-photon scattering amplitude. The difference between these amplitudes in the latter model and the minimal model are expressed by terms from the minimal model in a fairly compact fashion. The outcome confirms that these extra terms can not be suppressed by extra parameters in the second model except at the trivial limits.

We consider the identity found for the minimal model before and after reversible SW map as the manifestation of the formal equivalence between on shell effective actions of these two theories found in an earlier works~\cite{Martin:2016hji,Martin:2016saw}. While we focus on left-charged fermions in this work, the method could be generalized in a straightforward manner to right-charged and adjoint fermions in principle. We are therefore confident that the identity is a property of all NCQED models based on reversible SW maps. Once the SW map becomes irreversible the formal equivalence breaks down consequently. And as we have seen in Sec. IV the scattering amplitudes of the second model becomes different from the first model exactly via the deviations of the cancellation of the SW map contributions. One particular part of the two-by-two scattering amplitudes in second model, namely the collinear divergences, still follow those of the minimal model. Taking into account the nontrivial nature of the collinear divergences of the two-by-two scattering amplitudes of the minimal model itself~\cite{Latas:2020nji}, it could be worthy to study the IR behaviors of the scattering amplitudes of both the first and the second models in the future.

\acknowledgments{
We are grateful to C.P. Martin and P. Schupp for many discussions regarding SW maps.  J.T. would also like to acknowledge support of D. L\"ust and MPI,  M\"unchen,  for hospitality. 
}

\end{document}